\documentclass[fleqn,twoside]{article} 
\usepackage{epsf,multicol}
\usepackage{ujp}
\usepackage[cp866nav]{inputenc}
\usepackage[english,ukrainian]{babel}
\usepackage{mathrsfs}

\usepackage{amsmath}

\usepackage{latexsym}
\usepackage{amssymb}
\usepackage{amscd}
\usepackage{amsfonts}
\usepackage{stmaryrd}
\usepackage{eufrak}
\usepackage{bbold}

\newcommand{\beqs}{\begin{equation*}}
\newcommand{\beq}{\begin{equation}}

\newcommand{\eeqs}{\end{equation*}}
\newcommand{\eeq}{\end{equation}}

\newcommand{\beqas}{\begin{eqnarray*}}
\newcommand{\beqa}{\begin{eqnarray}}

\newcommand{\eeqas}{\end{eqnarray*}}
\newcommand{\eeqa}{\end{eqnarray}}

\newcommand{\seq}[5]{\newline 
\parbox{#1}{\begin{eqnarray*} #2 \end{eqnarray*}} \hfill
\parbox{#3}{\begin{eqnarray*} #4 \end{eqnarray*}} \hfill
\parbox{1cm}{\begin{equation} \label{#5} \end{equation}} 
\newline}

\newcommand{\eq}[2]{\begin{equation} #1 \label{#2} \end{equation}}

\newcommand{\meq}[2]{\begin{multline} #1 \label{#2} \end{multline}}

\newcommand{\eps}{\varepsilon}

\newcommand{\om}{\omega}
\newcommand{\ka}{\kappa}

\newcommand{\si}{\sigma}

\newcommand{\blist}{\begin{itemize}}

\newcommand{\elist}{\end{itemize}}

\providecommand{\href}[2]{#2}

\udk{04.60.Kz; 04.60.Gw;}

\udk{ 11.10.Lm; 11.80.Et}

\nazva{
A NOTE ON THE TRIVIALITY OF $\boldsymbol{\kappa}$-DEFORMATIONS OF GRAVITY
}

\nazvacol{A NOTE ON THE TRIVIALITY OF $\kappa$-DEFORMATIONS}%

\avtor{D.~GRUMILLER, W.~KUMMER, D.V.~VASSILEVICH~$^1$}%

\avtorcol{D.~GRUMILLER, W.~KUMMER, D.V.~VASSILEVICH}%

\inst{
Theoretische Physik, Technische Universit\"at Wien
}

\insti{
Intitut f\"{u}r Theoretische Physik, Universit\"{a}t Leipzig
}

\adr{(Wiedner Hauptstr.  8--10, A-1040 Wien, Austria)}

\adri{(Augustusplatz 10, D-04109 Leipzig, Germany)}

\begin{document}           

\maketitl                 

\begin{multicols}{2}
\anot{%
Some simple observations concerning certain $\ka$-deformations in the 
framework of $2D$ dilaton gravity formulated as Poisson-sigma models 
are discussed, and the question of (non-)triviality is addressed. 
}

\section{Introduction}

 Certain astrophysical observations have motivated the study of 
$\ka$-deformed Poincar{\'e}-algebras, e.g.\ in the context of ``Doubly 
Special Relativity'' (henceforth DSR; for a recent review and more references cf.\ 
e.g.\ \cite{Amelino-Camelia:2002vy,Magueijo:2002am,Magueijo:2001cr}). 
It might be helpful to consider the simpler case of $D=2$ first to 
settle open conceptual questions (especially concerning gravity). The 
first work in this direction appeared only recently 
\cite{Mignemi:2002hd}. In the present note when talking about 
$\ka$-deformations\footnote{Of course, $\ka$-deformations do not automatically imply DSR, nor vice versa.} we will exclusively refer to the deformation 
introduced by Magueijo and Smolin \cite{Magueijo:2001cr}, which differs 
from the original proposal \cite{Lukierski:1991pn}.\footnote{By $\ka$-deformation of Poincar{\'e} algebra only the deformation of the algebraic sector is considered in the present work, characterized by some fundamental mass parameter $\ka$. In \cite{Lukierski:1991pn} there was proposed the framework with quantum $\ka$ deformation, introducing as well the deformed nonsymmetric coproducts in the
coalgebra sector, which leads to noncommutative space-time \cite{Majid:1994cy}.}

 Moreover, it seems to be an interesting task by itself for purely 
mathematical reasons. Indeed, Izawa has shown a few years ago, that the 
most general consistent deformation (in the sense of Barnich and 
Henneaux \cite{Barnich:1993vg}) of a Poisson-$\si$ model (PSM) is again 
a PSM with the same number of target space coordinates 
\cite{Izawa:1999ib}. Since dilaton gravity without matter in $2D$ is 
merely a very special PSM \cite{Schaller:1994es} this result applies to 
it as well. Thus, generic consistent deformations of dilaton gravity in 
$2D$ are mathematically feasible.

 The description of dilaton gravity in terms of a PSM has turned out to 
be very fruitful -- e.g. all classical solutions have been obtained 
locally and globally within this approach \cite{Klosch:1996fi} and even 
in the presence of matter an exact quantization of geometry has been 
achieved \cite{Kummer:1997hy}, with interesting phenomenological 
applications for scattering processes \cite{Grumiller:2000ah}. For a 
recent review on dilaton gravity in $2D$ ref.\ \cite{Grumiller:2002nm} 
can be consulted.

 In the present note we intend to discuss a very simple question in the 
framework of twodimensional dilaton gravity: are $\ka$-deformations 
trivial or not (and in what sense are they (non)trivial)?

\section{Triviality of deformations}

 Let the momenta $p_a$ and the generator of boosts $J$ satisfy the 
undeformed Poincar{\'e} algebra in $2D$:  
\begin{equation} 
[p_0,p_1]=0,\quad [J,p_0]=p_1,\quad [J,p_1]=p_0\,.
\label{undef}
\end{equation}
An elementary calculation shows that the generators
\begin{equation}
P_a=\frac{p_a}{1+p_0/\kappa} \,,\quad J\,,
\label{pP}
\end{equation}
with $a=0,1$ satisfy the $\kappa$-deformed relations:
\seq{2cm}{
&& \left[P_a,P_b\right]=0\,,\\
&& \left[J,J\right]=0\,,
}{4cm}{
&& \left[J,P_0\right]=P_1-\kappa^{-1} P_1P_0 \,,\\
&& \left[J,P_1\right]=P_0-\kappa^{-1}P_1^2 \,.
}{deformed}
 Obviously for $\ka\to\infty$ the undeformed algebra is recovered. It 
is also instructive to invert (\ref{pP}):
\begin{equation}
p_a=\frac{P_a}{1-P_0/\kappa} \,.\label{Pp1}
\end{equation}
 Clearly the deformation is algebraically trivial.\footnote{This has 
been pointed out in ref.\ \cite{Magueijo:2002am}. This statement also 
reminds one of the rigidity results on deformations of Lie algebras 
\cite{Gerstenhaber:1964}.} The whole effect is a change of variables 
(\ref{pP}). Therefore, it seems natural to do the {\it same} change of 
variables in the PSM, the action of which reads \cite{Schaller:1994es}

\eq{
L_{\rm PSM} = \int_{M_2} \left[dX^I\wedge A_I + 
\frac{1}{2} \mathcal{P}^{IJ}A_J\wedge A_I\right]\,.
}{eq:PSM}
 For dilaton gravity (assuming Lorentzian signature for definiteness) 
the three gauge field 1-forms are $A_I=(\om,e^-,e^+)$, where $e^\pm$ is 
the dual basis of 1-forms in light-cone gauge for the anholonomic frame 
and $\om$ is related to the spin-connection 1-form via 
$\om^a{}_b=\eps^a{}_b\om$. The three target space coordinates 
$X^I=(X,X^+,X^-)$ contain the so-called ``dilaton'' $X$ and two 
auxiliary fields $X^\pm=(X^0\pm X^1)$ (again in light-cone 
representation). The Poisson-tensor ${\mathcal P}^{IJ}$ depends solely 
on the target space coordinates, it is antisymmetric and fulfills the 
Jacobi-identity ${\mathcal P}^{IN}\partial_N {\mathcal P}^{JK}+{\rm 
cycl.} (IJK)=0$. Due to the odd number of target space coordinates this 
tensor cannot have full rank and hence for non-trivial models exactly 
one Casimir function exists, related to the ``energy'' of the spacetime 
\cite{Grosse:1992vc}.

 The simplest dilaton gravity model is obtained with the ``free'' 
Poisson tensor
\begin{equation}
\mathcal{P}=\left( \begin{array}{ccc} 0&0&-X^1\\
0&0&-X^0\\ X^1&X^0&0 \end{array} \right)\,,
\label{Pfree}
\end{equation}
yielding vanishing curvature and torsion. In primed coordinates 
\begin{equation}
X^{a'}=\frac{X^a}{1+X^0/\kappa} \,,\qquad X'=X
\label{chX}
\end{equation} 
the Poisson tensor transforms as
\begin{equation}
\mathcal{P}^{I'J'}=\frac{\partial X^{I'}}{\partial X^I}
\frac{\partial X^{J'}}{\partial X^J}
\mathcal{P}^{IJ}\,. \label{calP}
\end{equation}
Consequently, in the primed coordinates we obtain again
the free PSM with an extra term,
\begin{equation}
\kappa^{-1} X^{1'}X^{a'} \omega \wedge e^{a'}\,,
\label{Mignemi}
\end{equation}
which is the result of Mignemi \cite{Mignemi:2002hd}.

 It seems natural to deform other dilaton models (spherically reduced 
gravity, the CGHS model \cite{Callan:1992rs}, the JT model 
\cite{Barbashov:1979bm}, etc.) in the same way \cite{Mignemi:2002tc} 
and to study the corresponding global structure. Thus, the starting 
point is the (undeformed) dilaton gravity action 
\begin{equation}
\hspace*{-0.8truecm}L = \int_{M_2} \left[ X_a De^a + X d\omega + \epsilon \left(X^+X^-U(X)+V(X)\right) \right]\,,
\label{2.62}
\end{equation}
with $\epsilon=e^+\wedge e^-$ being the volume 2-form and $X_a De^a = 
X^+ (d-\omega)\wedge e^- + X^- (d+\omega)\wedge e^+$ contains the torsion 2-form in light-cone representation. It is actually not necessary to 
restrict oneself to potentials of the type $X^+X^-U(X)+V(X)$ -- a 
particular class of relevant counter examples can be found in ref.\ 
\cite{Grumiller:2002md} -- but for sake of simplicity this special form 
will be assumed in the present work because it covers all special 
models referred to above. The following equations will be useful
\begin{eqnarray}
&&\frac{\partial X^{0'}}{\partial X^0} =
\left( 1-\frac{X^{0'}}{\kappa} \right)^2 \,,\quad\frac{\partial X^{1'}}{\partial X^1} = 1-\frac{X^{0'}}{\kappa}  \,,\nonumber \\
&&\frac{\partial X^{1'}}{\partial X^0} =
-\frac {X^{1'}}{\kappa} \left( 1-\frac{X^{0'}}{\kappa} \right) \,.
\label{deri}
\end{eqnarray}
By employing (\ref{calP}), (\ref{deri}) together with the relation
\begin{equation}
X^+X^-=\frac{X^{+'}X^{-'}}{\left( 1-\frac{X^{0'}}{\kappa} \right)^2}\,,
\end{equation}
the transformation law for the potentials is established:
\begin{equation}
\hspace*{-0.8truecm}U\to U'=U\left( 1-\frac{X^{0'}}{\kappa} \right)\,,\quad
V\to V'=V\left( 1-\frac{X^{0'}}{\kappa} \right)^3
\label{newUV}
\end{equation}

 Of course, the change of variables considered above is not an 
equivalence transformation: (i) it is singular; (ii) it changes the 
definition of the metric.

 Invariants of the deformed model can be traced back easily to the 
undeformed case. For example, the ``line element'' 
\meq{
\hspace*{-0.2truecm}(ds)^2=e^{0'}\otimes e^{0'} \left( 1-\frac{\eta_0}\kappa \right)^4
-\left(e^{1'}\otimes e^{0'}+e^{0'}\otimes e^{1'}\right) \frac {\eta_1}{\kappa} 
\times \\ 
\hspace*{-1truecm}\times \left( 1-\frac{\eta_0}\kappa \right)^3 -e^{1'}\otimes e^{1'} 
\left( 1-\frac{\eta_0}\kappa \right)^2 
\left( 1-\frac{\eta_1^2}{\kappa^2} \right)
}{dsprime}
 with $\eta_a:=X^{a'}$ results from a target space diffeomorphism 
(\ref{calP}) of the PSM, applied to the standard $(ds)^2$ and therefore 
must be invariant under the deformed transformations. Clearly, a 
description of the deformed geometry by means of (\ref{dsprime}) does 
not make much sense because the deformation would be without effect on 
classical solutions since all singularities in $e^{0'}, e^{1'}$ are 
compensated by corresponding zeros. This follows already from the 
previous general considerations, but it can be checked explicitly by 
plugging the explicit solutions (\ref{defe}) below into 
(\ref{dsprime}). The result is the original line element 
$(ds)^2=e^0\otimes e^0-e^1\otimes e^1$.

 On the other hand, taking a non-invariant object as ``metric'' seems 
to be questionable from a physical point of view. For instance, the 
curvature scalar related to the metric presented in 
\cite{Mignemi:2002hd,Mignemi:2002tc} depends on the choice of gauge, 
which from a relativistic point of view is an inconvenient feature.

 Thus there seems to be no way to evade both, the Scylla of triviality 
and the Charybdis of non-invariance.

\section{Deformed solutions}

 For a more explicit analysis we have to express the deformed 
``zweibein'' in terms of the undeformed variables,
\begin{eqnarray}
&&e^{0'}=\left( 1+\frac{X^0}\kappa \right)^2 e^0 +
\frac{X^1}\kappa \left( 1+\frac{X^0}\kappa \right) e^1\,,\nonumber \\
&&e^{1'}=\left( 1+\frac{X^0}\kappa \right) e^1 \,.\label{defe}
\end{eqnarray}
 If we require non-triviality of the deformation a non-invariant 
``metric'' has to be employed, as discussed in the previous section. 
Since there does not seem to exist a better alternative (there is no 
canonical choice), let us assume that the ``physical metric'' is 
constructed in the usual way from the deformed ``zweibein'' 
(\ref{defe}), namely $g_{\mu\nu}'=e^{a'}_\mu\otimes 
e^{b'}_\nu\eta_{ab}$, where $\eta_{ab}$ is the flat 
metric.\footnote{The words ``metric'' and ``zweibein'' have been put 
under quotation marks to indicate that this nomenclature is too 
suggestive. Neither $e^{a'}$ nor $g_{\mu\nu}'$ transform as the 
notation seems to promise (unless $\ka$ is taken to $\infty$).} 

 As a demonstration one can choose spherically reduced gravity (from 
arbitrary dimension $D$) as undeformed starting point. Equations of 
motion imply
\begin{equation}
X^a =\epsilon^{\mu\nu} e_\mu^a \partial_\nu X\,.
\label{R2.30}
\end{equation}

 The discussion will be restricted to the behavior of the deformed 
solution in the asymptotic region of the undeformed one. This is the 
simplest test for the validity of the deformed model. In this region we 
can apply Schwarzschild gauge $(ds)^2=\xi(dt)^2-\xi^{-1}(dr)^2$ with 
$\xi(r)$ and $X(r)$ defined by \cite{Katanaev:1997ni} 
\begin{equation}
\label{e23EF}
\xi \left( r\right) =
1+2\mathcal{C}_{0}\left| 1-a\right|^{\frac{a}{a-1}}
r^{\frac{a}{a-1}}\left( \frac{B}{a}\right)^{\frac{2-a}{2
\left( a-1\right) }}\,,
\end{equation}
and 
\begin{equation}
\label{e23ur}
r=\sqrt{\frac{a}{B}}\frac{1}{|1-a|}X^{1-a}\,,
\end{equation}
 respectively. The constants in these equations have the following 
meaning: $a=(D-3)/(D-2)$, $B$ is a normalization constant (which can be 
set to 1) and ${\mathcal C}_0$ is the value of the Casimir function 
(for positive/negative values a naked singularity/black hole is 
described, respectively; if it vanishes the Minkowskian ground state is 
reached). The (singular) limit $D\to\infty$, if treated with care, 
yields the CGHS model. 

Asymptotically ($r\to \infty$) one readily obtains
\begin{equation}
\hspace*{-0.5truecm}X\sim r^{1/(1-a)},\qquad X^1=0,\qquad X^0\sim r^{a/(1-a)}.
\label{Xr}
\end{equation}
One observes that for $\frac 12 \le a <1$ (i.e. for spherical 
reduction with $4\leq D<\infty$) the function
\begin{equation}
\chi (r)=1+\frac{X^0}\kappa \label{chir}
\end{equation}
 which characterizes the ``strength'' of the deformation is large, 
implying that asymptotically the model is being deformed 
noticeably.\footnote{It is somewhat amusing that $D=4$ is the limiting 
case.} This indicates that we are dealing in fact with the old Fock 
version \cite{Fock:1964} of the deformations which modifies physics at 
large distances. Indeed, the way in which the generators $X^I$ have 
been interpreted corresponds to a deformation in coordinate space 
rather than in momentum space: for instance, the dilaton $X$ has been 
regarded as some (power of) a ``radius'' and thus as a coordinate space 
entity. However, these interpretational issues are not pivotal for the 
present discussion.

 Despite of this sizable asymptotic deformation the scalar curvature 
for the primed zweibein reads
\begin{equation}
R(r)\sim r^{-2/(1-a)}\,. \label{Rprime}
\end{equation}
 Thus, the asymptotically flat region remains asymptotically flat after 
the deformation, which is an attractive feature and indicates that it 
may be sensible to talk of an ``asymptotic region'' even in the 
deformed case.

 It should be emphasized strongly, though, that one should not 
over-interpret results extracted from a ``metric'' which does not 
possess the (deformed) Lorentz invariance.

\section{Conclusion}

 We have shown that $\ka$-deformed dilaton gravity in $2D$ is either 
classically equivalent to a corresponding undeformed model (and thus 
the deformation would be trivial) or one has to deal with a 
``non-invariant metric''.

 However, this does not imply that all deformations are without effect. 
In particular, the question of the ``correct'' metric turns out to be a 
non-trivial one, even after imposing the invariance condition. This 
will be the subject of work in progress dealing with these issues in a 
much more comprehensive manner \cite{ts,gkmsv}.

We stress that only the classical part of the full 
$\kappa$-deformed Poincar{\'e} bialgebra is being used in our
construction. Our procedure may be as well called ``deformed gravity with an invariant energy scale''. An interesting 
development  may consist in a combination of the 
Poisson-$\si$ models with the quantum algebra approach to DSR
described in the recent papers \cite{Kowalski-Glikman:2002jr}.

 It also would be worthwhile to investigate to which extent our 
conclusions depend on a particular choice of the basis in the 
$\kappa$-deformed Poincar{\'e} algebra.

 During the final preparations of this proceedings contribution an 
e-print appeared which has partial overlap with our discussion 
\cite{Ahluwalia:2002wf}. The authors of that paper conclude that DSR is operationally indistinguishable from special relativity.

\section*{Acknowledgement}

 We would like to thank our collaborators on deformations of dilaton 
gravity in $2D$, S.\ Mignemi and T.\ Strobl, for enlightening 
discussions and helpful correspondence. Moreover, we are indebted to 
V.\ Lyakhovsky for his comments on deformations of algebras. 
We thank G.\ Amelino-Camelia for his interest and correspondence. Finally, we are grateful to J.\ Lukierski for relevant critical remarks.

 One of the authors (D.G.), who has presented the results of this note 
at the XIV~$^{\rm th}$ International Hutsulian Workshop on 
``Mathematical Theories and their Physical \& Technical Applications'' 
in Cernivtsi (Ukraine), is grateful to the organizers and participants 
of this meeting for their hospitality and numerous interesting 
conversations, in particular L.\ Adamska, S.\ Moskaliuk, S.\ 
Rumyantsev, A.\ Vlassov and M.\ Wohlgenannt. 

 This work has been supported by project P-14650-TPH of the Austrian 
Science Foundation (FWF), by DFG project BO 1112/12-1, by the MPI MIS 
(Leipzig) and by the Austro-Ukrainian Institute for Science and 
Technology.




\begin{thebibliography}{99}

\bibitem{Amelino-Camelia:2002vy}
{\it Amelino-Camelia G.} Doubly-special relativity: First results 
and key open problems // arXiv.org/abs/gr-qc/0210063.  

\bibitem{Magueijo:2002am}
{\it Magueijo~J. and Smolin~ L.} Generalized Lorentz invariance 
with an invariant energy scale //  
http: // arXiv.org/abs/gr-qc/0207085.

\bibitem{Magueijo:2001cr}
{\it Magueijo~J. and Smolin~L.} Lorentz invariance with an invariant 
energy scale //  Phys. Rev. Lett. -- 2002. -- {\bf 88}. -- 190403  // 
arXiv.org/abs/hep-th/0112090.

\bibitem{Mignemi:2002hd}
{\it Mignemi~S.} Two-dimensional gravity with an invariant energy 
scale //  arXiv.org/abs/hep-th/0208062.  

\bibitem{Lukierski:1991pn}
{\it Lukierski~J., Ruegg~H., Nowicki~A. and Tolstoi~ V.~N.} 
$Q$-deformation of Poincar\'e algebra //  Phys.\ Lett. -- 1991. -- {\bf 
B264}. -- 331.  

\bibitem{Majid:1994cy}
{\it S.~Majid and H.~Ruegg} Bicrossproduct structure of kappa Poincar\'e group and noncommutative geometry //  Phys.\ Lett. -- 1994. -- {\bf B334}. --  348  // 
arXiv.org/abs/hep-th/9405107.
\\
{\it J.~Lukierski, H.~Ruegg and W.~J.~Zakrzewski} Classical quantum mechanics of free kappa relativistic systems //  Annals Phys. -- 1995. -- {\bf 243}. -- 90  //  arXiv.org/abs/hep-th/9312153.

\bibitem{Barnich:1993vg}
{\it Barnich~G. and Henneaux~M.} 
Consistent couplings between fields with a gauge
freedom and deformations of the master equation //  Phys. Lett. -- 1993. 
 -- {\bf B311}. -- 123--129  // 
arXiv.org/abs/hep-th/9304057; \\
{\it Barnich~G., Brandt~F., and Henneaux~M.} Local BRST cohomology 
in gauge theories //  Phys. Rept. -- 2000. -- {\bf 338}. -- 439--569  // 
arXiv.org/abs/hep-th/0002245.  

\bibitem{Izawa:1999ib}
 {\it Izawa K.~I.} On nonlinear gauge theory from a deformation theory
  perspective //  Prog. Theor. Phys. -- 2000. -- {\bf 103}. -- 
225--228  // 
arXiv.org/abs/hep-th/9910133. 

\bibitem{Schaller:1994es}
{\it Schaller~P. and Strobl~T.} Poisson structure induced 
(topological) field theories //  Mod. Phys. Lett. -- 1994. -- {\bf A9}. 
 -- 3129--3136  // 
 arXiv.org/abs/hep-th/9405110.  

\bibitem{Klosch:1996fi}
{\it Kl\"osch~T. and Strobl~T.} Classical and quantum gravity in
  (1+1)-dimensions. Part 1: A unifying approach //  Class. Quant.
  Grav. -- 1996. -- {\bf 13}. -- 965--984  // 
arXiv.org/abs/gr-qc/9508020;\\
{\it Kl\"osch~T. and Strobl~T.} Classical and quantum gravity in 
1+1 dimensions.  Part II: The universal coverings //  Class. 
Quant.  Grav. -- 1996. -- {\bf 13}. -- 2395--2422  // 
arXiv.org/abs/gr-qc/9511081;\\
{\it Kl\"osch~T. and Strobl~T.} Classical and quantum gravity in 1+1
  dimensions. Part III: Solutions of arbitrary topology and kinks in 1+1
  gravity //  Class. Quant. Grav. -- 1997. -- {\bf 14}. -- 1689--1723  // 
arXiv.org/abs/hep-th/9607226;\\ 
{\it Kl\"osch~T. and Strobl~T.}
A global view of kinks in 1+1 gravity //  Phys.\ Rev. -- 1998. -- {\bf 
D57}. -- 1034  // 
arXiv.org/abs/gr-qc/9707053;\\ 
{\it Strobl~T.} Gravity in two spacetime dimensions // 
arXiv.org/abs/hep-th/0011240.

\bibitem{Kummer:1997hy}
 {\it Kummer~W., Liebl~H., and  Vassilevich~D.~V.} Exact path 
integral quantization of generic $2-d$ dilaton gravity //  Nucl. 
Phys. -- 1997. -- {\bf B493}. -- 491--502  // 
arXiv.org/abs/gr-qc/9612012;\\ 
Integrating geometry in general $2d$ dilaton gravity with matter //  
 Nucl. Phys. 1999. -- {\bf B544}. -- 403--431  //  
arXiv.org/abs/hep-th/9809168; 
{\it Grumiller~D.} Quantum dilaton gravity in two dimensions with 
matter.  //   PhD thesis, Technische Universit\"at Wien, 2001  // 
arXiv.org/abs/gr-qc/0105078.  

\bibitem{Grumiller:2000ah}
{\it Grumiller~ D., Kummer~W., and  Vassilevich~D.~V.} The virtual 
black hole in $2d$ quantum gravity //  Nucl. Phys. -- 2002. -- {\bf 
B580}. -- 438--456  // 
arXiv.org/abs/gr-qc/0001038;\\ 
{\it Fischer~P., Grumiller~D., Kummer~W., and  Vassilevich~D.~V.} 
$S$-matrix for $s$-wave gravitational scattering //  Phys. Lett. -- 
2001. -- {\bf B521}. -- 357--363  // 
  arXiv.org/abs/gr-qc/0105034.\\  
  Erratum ibid. -- 2002. -- {\bf B532}. -- 373; \\  
{\it Grumiller~D.} The virtual black hole in $2D$ quantum gravity 
and its relevance for the $S$-matrix //  Int. J.  Mod. Phys. -- 2001. 
  -- {\bf A 17}. -- 989  // 
arXiv.org/abs/hep-th/0111138;\\ 
Virtual black hole phenomenology 
from $2d$ dilaton theories //  Class. Quant. Grav. -- 2002. -- {\bf 19}. 
  -- 997--1009  // 
arXiv.org/abs/gr-qc/0111097;\\ 
{\it Grumiller~D., Kummer~W., and Vassilevich~D.V.} 
  Virtual black 
holes in generalized dilaton theories and their special role in string 
  gravity  //  arXiv.org/abs/hep-th/0208052.  

\bibitem{Grumiller:2002nm}
{\it Grumiller~D., Kummer~W., and Vassilevich~D.V.} 
  Dilaton gravity in two  dimensions //  Phys. Rept. -- 2002. -- {\bf 
  369}. -- 327--429  // 
arXiv.org/abs/hep-th/0204253.  

\bibitem{Gerstenhaber:1964}
 {\it Gerstenhaber~M.} On the deformation of rings and algebras // 
Ann. Math. -- 1964. -- {\bf 79}. -- 59;\\ 
{\it Levy-Nahas~M.}, 
Deformation and contraction of Lie algebras //  J. Math.  
Phys. -- 1967. -- {\bf 8}. -- 1211;\\
{\it Lyakhovsky V.~D.} Generalized symmetries and deformations of 
  the direct sums of Lie algebras //  Commun.  Math. Phys. -- 1968. -- 
  {\bf 11}. -- 131.

\bibitem{Grosse:1992vc}
{\it Grosse~H., Kummer~W., Presnajder~P., and Schwarz~D.~J.}
  Novel symmetry of nonEinsteinian gravity in two- dimensions // 
  J.  Math. Phys. -- 1992. -- {\bf 33}. -- 3892--3900  // 
arXiv.org/abs/hep-th/9205071;\\
{\it Mann~ R.~B., } Conservation laws and $2-d$ black holes in dilaton 
  gravity //  Phys. Rev. -- 1993. -- {\bf D47}. -- 4438--4442  // 
arXiv.org/abs/hep-th/9206044;\\
{\it Kummer W., and Widerin P.} Non-Einsteinian gravity in $d=2$: 
Symmetry and current algebra //  Mod. Phys. Lett. -- 1994. -- {\bf A9}. 
-- 1407--1414; \\
Conserved quasilocal quantities and general covariant theories in 
  two-dimensions //  Phys. Rev. -- 1995. -- {\bf D52}. -- 6965--6975  //  
arXiv.org/abs/gr-qc/9502031;\\
{\it Kummer W., and Lau S.~R.} Boundary conditions and quasilocal 
energy in the canonical formulation of all 1 + 1 models of gravity //  
 Annals Phys. -- 1997. -- {\bf 258}. -- 37--80  // 
arXiv.org/abs/gr-qc/9612021;\\
{\it Kummer W. and Tieber G.} Universal conservation law and 
 modified N\"other symmetry in $2d$ models of gravity with matter // 
 Phys.  Rev. -- 1999. -- {\bf D59}. -- 044001  // 
arXiv.org/abs/hep-th/9807122;\\
{\it Grumiller D.  and Kummer W.} Absolute conservation law for 
black holes //  Phys. Rev. -- 2000. -- {\bf D61}. -- 064006  // 
arXiv.org/abs/gr-qc/9902074.

\bibitem{Callan:1992rs}
{\it Callan C.~G.(Jr.), Giddings S.~B.,  Harvey J.~A., and 
Strominger A.} Evanescent black holes //  Phys. Rev. -- 1992. -- {\bf 
  D45}. -- 1005--1009  // 
arXiv.org/abs/hep-th/9111056.

\bibitem{Barbashov:1979bm}
 {\it Barbashov B.~M.,  Nesterenko V.~V., and  Chervyakov A.~M.} 
  The solitons in some  geometrical field theories //  Theor. Math. 
Phys. -- 1979. -- {\bf 40}. -- 572--581 (Teor. Mat. Fiz. -- 1979.  -- 
{\bf 40}.--  15--27), J.  Phys.  -- 1980. -- {\bf  A13}. -- 
301--312; \\
{\it D'Hoker E.,  and Jackiw R.} 
Liouville field theory //  Phys. Rev. -- 1982. --  {\bf D26}. -- 3517; \\
{\it Teitelboim C.} Gravitation and 
Hamiltonian structure in two space-time dimensions //  Phys.  
Lett. -- 1983. -- {\bf B126}. -- 41; \\
{\it  D'Hoker E., Freedman D., and Jackiw R.} $SO(2,1)$ invariant 
quantization of the Liouville theory //  Phys. Rev. -- 1983. -- {\bf 
D28}. --2583; \\
{\it D'Hoker E., and Jackiw R.} Space translation breaking and 
compactification in the Liouville theory //  Phys.  Rev. Lett. --
1983. -- {\bf 50}. -- 1719--1722; \\  
{\it Jackiw R.} Another view on massless matter-gravity fields in 
two-dimensions //  arXiv.org/abs/hep-th/9501016.  

\bibitem{Mignemi:2002tc}
{\it Mignemi S.} Two-dimensional gravity with an invariant energy 
scale and arbitrary dilaton potential //  
arXiv.org/abs/hep-th/0210213.

\bibitem{Grumiller:2002md}
{\it Grumiller D. and Vassilevich D.~V.} Non-existence of a dilaton 
gravity action for the exact string black hole //  JHEP. -- 2002. -- {\bf 
  11}. --  018  // 
arXiv.org/abs/hep-th/0210060.

\bibitem{Katanaev:1997ni}
{\it  Katanaev M.~O., Kummer W., and Liebl H.} On the completeness 
of the black hole singularity in $2d$ dilaton theories //  Nucl. 
Phys. -- 1997. -- {\bf B486}. -- 353--370  // 
arXiv.org/abs/gr-qc/9602040.

\bibitem{Fock:1964}
{\it  Fock V.~A.} The theory of space, time and gravitation. --
Oxford: Pergamon Press, 1964;\\
{\it  Manida S.~N.} Fock-Lorentz transformations and time-varying 
speed of light  //  arXiv.org/abs/gr-qc/9905046.  

\bibitem{ts}
{\it Strobl T.} (in preparation).

\bibitem{gkmsv}
{\it Grumiller D., Kummer W., Mignemi S., Strobl T., and Vassilevich 
D.} Deformations of dilaton gravity in $2d$ (in preparation).

\bibitem{Kowalski-Glikman:2002jr}
{\it Kowalski-Glikman J., and Nowak S.} Non-commutative space-time 
of doubly special relativity theories //  
arXiv.org/abs/hep-th/0204245;\\
{\it Lukierski J.,  and Nowicki A.} Nonlinear and quantum origin of 
doubly infinite family of modified addition laws for fourmomenta //  
arXiv.org/abs/hep-th/0209017; \\
{\it Kowalski-Glikman J.} Doubly special relativity: A kinematics of 
quantum gravity? // arXiv.org/abs/hep-th/0209264;\\ 
{\it Lukierski J.,  and Nowicki A.} Four 
classes of modified relativistic symmetry transformations // 
arXiv.org/abs/hep-th/0210111.

\bibitem{Ahluwalia:2002wf}
{\it Ahluwalia D.~V., Kirchbach M., and Dadhich N.} Operational 
indistinguishabilty of doubly special relativities from special 
relativity  // arXiv.org/abs/gr-qc/0212128.  


\begin{flushright}
{\footnotesize Received 1.13.2003}\\
{\footnotesize Accepted 3.3.2003}
\end{flushright}

\end{thebibliography}



\rezume 
{ЗАУВАЖЕННЯ  ПРО ТРИВЎАЛЬНЎСТЬ $k$-ДЕФОРМАЦЎЙ ГРАВЎТАЦЎ°}
{Д. Грумўлер, В. Кумер, Д.В. Васўлевўч}
{
В рамках $2D$ дилатонно∙ гравўтацў∙, сформульовано∙ в термўнах 
пуасонўвсько∙ сигма-моделў, обговорюються простў спостереження 
стосовно певних $k$-деформацўй, увага зосереджуїться наколо (не-) 
тривўальностў цих деформацўй.}

\rezume{
ЗАМЕЧАНИЕ О ТРИВИАЛЬНОСТИ $k$-ДЕФОРМАЦИЙ ГРАВИТАЦИИ
}
{ Д. Грумиллер, В. Куммер, Д.В. Василевич} 
{
В рамках $2D$ дилатонной гравитации, сформулированной в терминах 
пуассоновской сигма-модели, обсуждаются простые наблюдения 
относительно определенных $k$-деформаций, вопрос сосредоточен вокруг 
(не-) тривиальности этих деформаций.}

\end{multicols}
\end{document}